\documentclass[aps,pra,epsfigure,twocolumn,showpacs]{revtex4}
\usepackage{amsmath}
\usepackage{amstext}
\usepackage{latexsym}
\usepackage{graphicx}
\usepackage{amsfonts}
\usepackage{epsfig}
\usepackage{color}

\newcommand{\petr}[1]{\textcolor{black}{#1}} 

\begin{document}
\title{Nonclassicality in phase space and nonclassical correlation}

\author{Petr Marek}

\affiliation{Department of Optics, Palack\'{y} University, 17. listopadu 50, 77200 Olomouc, Czech Republic}
\affiliation{School of Mathematics and Physics, The Queen's University, Belfast BT7 1NN, United Kingdom}

\author{M. S. Kim}

\affiliation{School of Mathematics and Physics, The Queen's University, Belfast BT7 1NN, United Kingdom}

\author{Jinhyoung Lee}

\affiliation{Department of Physics, Hanyang University, Seoul 133-791, Korea}

\begin{abstract}
Continuous variable entanglement is a manifestation of
nonclassicality of quantum states. In this paper we attempt to
analyze whether and under which conditions nonclassicality can be
used as an entanglement criterion. We adopt the well-accepted
definition of nonclassicality in the form of lack of well-defined
positive Glauber Sudarshan P-function describing the state. After
demonstrating that the classicality of subsystems is not
sufficient for the nonclassicality of the overall state to be
identifiable with entanglement, we focus on Gaussian states and
find specific local unitary transformations required to arrive at
this equivalency. This is followed by the analysis of quantitative
relation between nonclassicality and entanglement.
\end{abstract}
\pacs{03.67.Mn,42.50.Dv,03.65.Wj}

\maketitle
\section{Introduction}
In continuous variables (CV) quantum optics, the nonclassicality of a light field is defined
using its quasiprobability distributions in phase space.  Because
of the uncertainty principle, it is not possible to have and exact
probability distribution in phase space description of a quantum
state; hence there is a family of quasiprobability distributions,
the most famous of which being the Husimi $Q$, the Wigner $W$ and
the Glauber-Sudarshan $P$ function~\cite{glauber}.  The
quasiprobability distribution of a system has one-to-one
correspondence to its density operator.   In this paper, we are
interested in the Glauber-Sudarshan $P$ function (for simplicity,
the $P$ function)~\cite{glauber-sudarshan}, which is a diagonal
representation of a state of density operator $\rho$ in the
coherent state basis $|\alpha\rangle$: $ \rho = \int d^2\alpha
P(\alpha) |\alpha\rangle\langle\alpha| ~~~; ~~~\alpha\in\mathbb{C}
$. Klauder formally proved that any state can be written in such
the sum as far as the $P$ function is not restricted to a regular
positive one~\cite{klauder}. In his seminal paper~\cite{mandel},
Mandel writes `{\it For the coherent state corresponds as closely
as possible to a classical state of definite complex amplitude.
When $P(\alpha)$ is a probability density, then the state
$\rho$ corresponds to an ensemble of different complex
amplitudes with ensemble density $P(\alpha)$, which is just how an
optical field is described classically.  But when $P(\alpha)$ is
not a probability density, the classical analogy fails completely,
and we have a purely quantum mechanical state.}' In extension of
this argument, we can write a two-mode classical state as
\begin{equation}
\rho=\int d^2\alpha d^2\beta P(\alpha,\beta)|\alpha\rangle\langle\alpha|\otimes|\beta\rangle\langle\beta|
\label{two}
\end{equation}
with a regular and positive two-mode $P$ function
$P(\alpha,\beta)$.

We can easily see that if $P(\alpha,\beta)$ is
a positive regular function which is normalized, the given field
is separable by definition; {\it a sufficient condition for
separability}.  However, although separability and nonclassicality are inherently linked \cite{Kim}, the existence of a positive $P$ function
is not necessarily a necessary condition for separability because
the negativity of the $P$ function may be from nonclassicality of
a local state rather than from the nonclassical correlation. For Gaussian states, this intricacy has been studied extensively \cite{duan,R5,R1,R2,R4,R3}. It was shown that entangled states are indeed only a subset of nonclassical states and that equivalency between these two properties can be found only for special cases with specific symmetry. Natural questions arise: does this symmetry need to be inherent to the respective quantum states? Can any quantum state be transformed (with help of entanglement preserving local unitary transformations) to a form for which entanglement and nonclassicality are equivalent?

Since classicality of a state is given by the nature of its $P$ function, one might intuitively look for analogies within the $P$ function of a global multi-mode state. For example, if a two-mode quantum state is nonclassical, while both its subsystems are classical, one might reason that this is a consequence of entanglement of the state. Therefore, if entanglement preserving local unitary transformations could be used to transform a state into a form with classical subsystems, nonclassicality of the global state would be an indicator of entanglement.  However, we are going to show that this approach is flawed, or rather incomplete.

The degree of entanglement does not change by local unitary
operations. Duan {\em et al.} utilize this property of
entanglement and the existence of the $P$ function to find out a
sufficient and necessary condition for separability~\cite{duan}.
However, nonclassicality, which was originally defined for a state
of single mode of light, does not possess this property -- local
squeezing, for example, can create a nonclassical state easily.
Therefore, when looking for a connection between nonclassicality
and entanglement, we have to consider a specific form to which
various states can be brought with help of local unitary
operations. Instead of using local unitary transformations to bring local subsystems to classical, we should in fact aim to reduce the global nonclassicality of the state as much as possible.
For Gaussian states we analyze the local unitary
transformations to find those which remove all the local
manifestations of nonclassicality.  We also explore
if there is a quantitative relation between nonclassicality and
entanglement.

\section{Global nonclassicality without entanglement}
Let us start by showing that classicality of local subsystems is by no means sufficient for global non-classicality to be taken as a sign of entanglement.
The state of a composite system is said to be
separable when its density operator $\rho$ is a convex sum:
$\rho=\sum_i p_i\rho_i^a\rho_i^b$, where $p_i$ is the probability
and $\rho_i^{a,b}$ are density operators of two subsystems $a$ and
$b$.

Consider a separable state of the following form:
\begin{equation}\label{state}
\rho = p \rho_{|\beta\rangle}^a\otimes \rho_{|1\rangle}^b + (1-p)
\rho_{|-\beta\rangle}^a\otimes\rho_{|0\rangle}^b,
\end{equation}
where $p$ is a probability $0\leq p\leq 1$ and $|\beta\rangle$,
$|1\rangle$ and $|0\rangle$ represent a coherent state, a single
photon state and the vacuum, respectively. The short notation of
density operator $\rho_{|\psi\rangle}$ denotes a state
$|\psi\rangle\langle\psi|$ with one extra unit of vacuum noise,
that is,
\begin{equation}
\rho_{|\psi\rangle} = \int \frac{2}{\pi} e^{-2|\alpha|^2} \hat{D}(\alpha)|\psi\rangle\langle\psi| \hat{D}(\alpha)^{\dag} d^2\alpha,
\end{equation}
where $\hat{D}(\alpha) = \exp(\alpha^* \hat{a} - \alpha
\hat{a}^{\dag})$ is the displacement operator. This ensures that
the $P$ function of any state under consideration is
regular~\cite{barnett-n}. It may still retain negative values
intrinsic to the initial pure state, as in the case of the
single-photon Fock state $|1\rangle$, but no longer does it show
singular behavior. With this in mind we can straightforwardly find
the $P$ function of the state (\ref{state}):
\begin{eqnarray}
P(\alpha_a,\alpha_b) &=& \frac{4p}{\pi}(4|\alpha_b|^2-1)e^{-2|\alpha_a-\beta|^2 - 2|\alpha_b|^2}
\nonumber \\
&&+ \frac{4(1-p)}{\pi}e^{-2|\alpha_a+\beta|^2-2|\alpha_b|^2}.
\end{eqnarray}
It can easily be seen that if the probability $p$ is suitably
chosen, this $P$ function for the whole system, although negative
at some points, may have its marginal distributions positive
everywhere in phase space.

\begin{figure}
\centerline{\psfig{figure=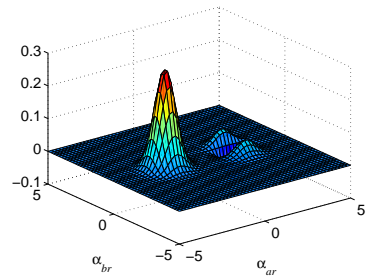,width=0.6\linewidth}}
\caption{(Color online) Cut of the $P$ function for state
(\ref{state}), $P(\alpha_a, \alpha_b)$,  along the plane,
$\alpha_{ai} = \alpha_{bi}  = 0$. \petr{The parameters employed are $\beta =
2$ and $p = 3/4$.} } \label{fig1}
\end{figure}

\begin{figure}[b]
  {\bf (a)}\hskip4.0cm{\bf (b)}
    \centerline{\includegraphics[width=0.2\textwidth]{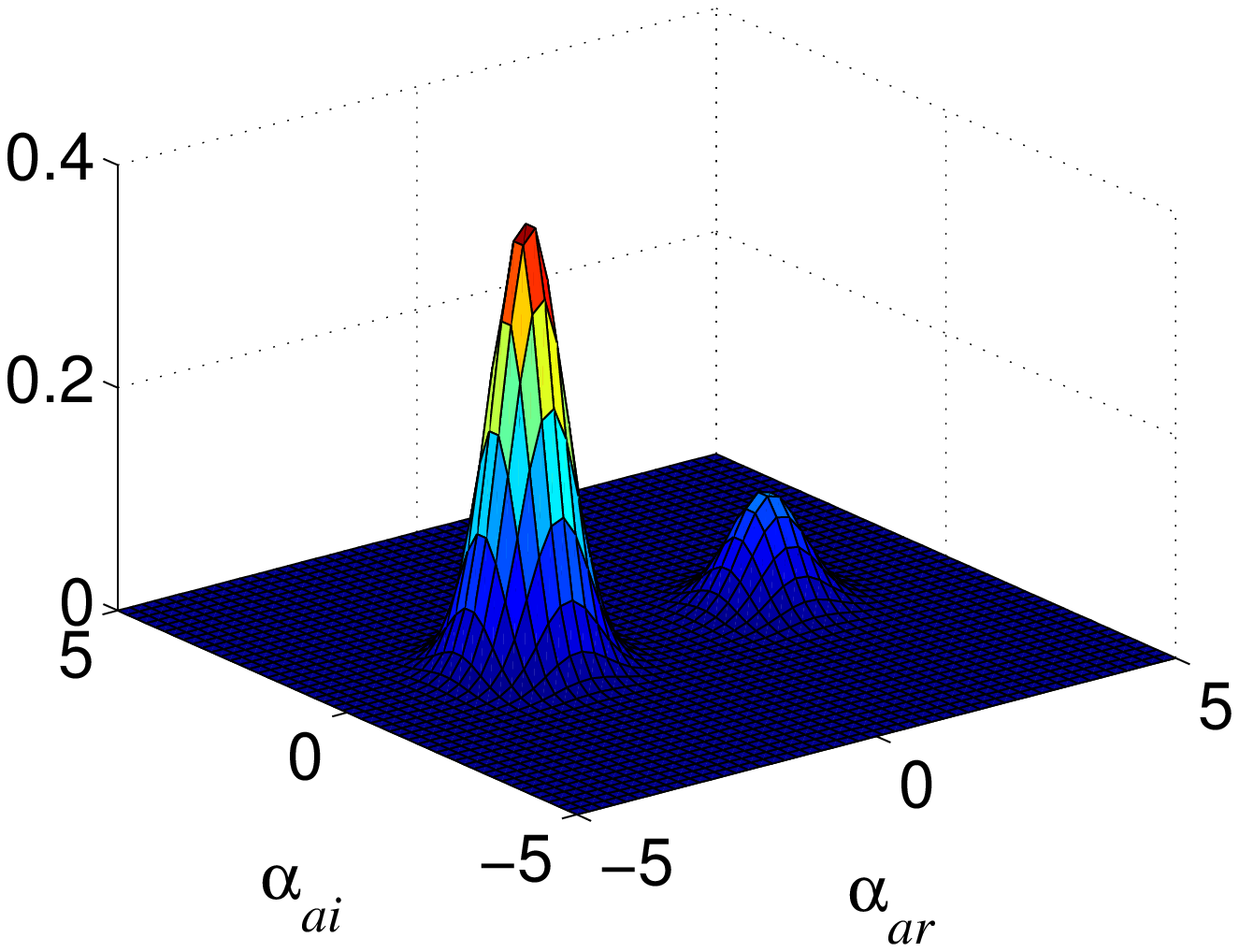}
    \hspace*{0.2cm}\includegraphics[width=0.2\textwidth]{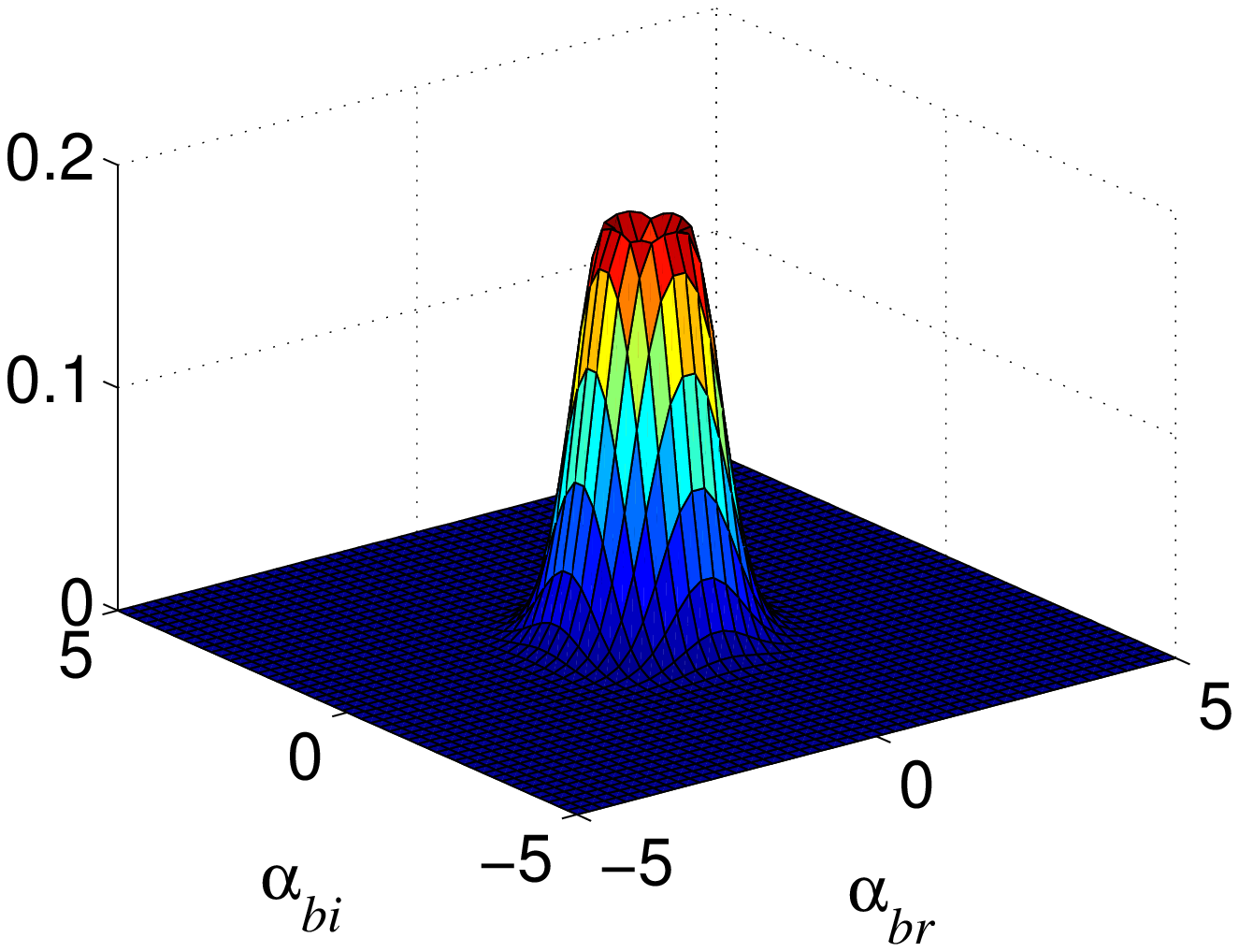}}
\caption{(Color online)  $P$ function for subsystem $a$ (in
{\bf(a)}) and subsystem $b$ (in {\bf (b)}) of state (\ref{state}).
$\beta=2$ and $p=3/4$.} \label{fig2}
\end{figure}

Let us consider state (\ref{state}) with the coherent amplitude
$\beta=2$ and the probability $p=3/4$. \petr{ Fig.~ \ref{fig1} shows a
two-dimensional cut $P(\alpha_{ar},\alpha_{br})$ of the four-dimensional $P$ function of the
state along the plane defined by $\alpha_{ai} = 0$ and
$\alpha_{bi} = 0$, where the subscripts `$i$' and `$r$' label the imaginary and the real
part of amplitudes $\alpha_{a,b}$, respectively.}  An area of negativity can
be seen in this cut. However, the $P$ functions for both the
subsystems, defined as
\begin{equation}
\rho^k=\mbox{Tr}_l[\rho]~~;~k, l=a, b~\mbox{and}~k\neq l,
\end{equation}
are positive everywhere in phase space as shown in Figs.
\ref{fig2}.  Now, we have a clear example of a composite system
whose  subsystems have positive and regular $P$ functions, whereas
the total $P$ function retains negative values.  The state is,
however, not entangled.  The non-classicality is due to
classical correlation.


It is obvious now that the classicality of subsystems is not the condition we are looking for. Instead, one should consider the whole range of local unitary operations and analyze whether any state can be transformed into a form for which there is equivalency between nonclassicality and entanglement. However, the difficulty of such the task is apparent at the first glance, because an arbitrary local unitary transformation has infinite number of parameters. It is therefore prudent to focus on Gaussian states, which are described by Gaussian Wigner functions and are fully defined by their first and second moments. Consequently, any Gaussian transformation can be readily described in a similar way.
The Gaussian states were, due to the mathematical handiness and experimental relevance, at the center of focus for CV entanglement studies and the sufficient and necessary condition of separability has been found ~\cite{duan,simon}. Several authors have also investigated the relation between nonclassicality and entanglement for Gaussian states \cite{R1,R2,R4} and found these two properties interchangeable only for a class of states which can be brought to a specific symmetric form, \petr{in which $|\langle \hat{x}_a\hat{x}_b\rangle| = |\hat{p}_a\hat{p}_b\rangle|$, $\langle \hat{x}_k^2\rangle = \langle \hat{p}_k^2\rangle$ for $k = a,b$, and the remaining moments are zero. Note that $\hat{x}_{a,b}$ and $\hat{p}_{a,b}$ are quadrature operators for modes $a$ and $b$, $\hat{a} = (\hat{x}_a + i\hat{p}_a)/\sqrt{2}$, obeying the commutation relation $[\hat{x},\hat{p}] = i$.} However, for all Gaussian states the existence of a positive $P$ function after appropriate local unitary operations is a necessary and sufficient condition \petr{for separability~\cite{lee}}. What remains is to explicitly specify the nature of these operations.

\section{Local operations to unveil Gaussian entanglement}
For Gaussian states with certain inherent symmetry, criteria for non-classicality and entanglement are equivalent \cite{R1,R2,R4}. However, only a subset of all Gaussian states can be brought to the required form. For general Gaussian states, different transformations are needed. \petr{Let us now consider local unitary operations with the goal of reducing global nonclassicality as much as possible, canceling it completely for separable states and leaving only entanglement otherwise.}

Duan {\it et al.}~\cite{duan} introduced a sufficient
condition for entanglement --  a two mode quantum state is
entangled, if it does not satisfy the following inequality:
\begin{equation}
\langle(\Delta \hat{u})^2\rangle + \langle(\Delta \hat{v})^2\rangle \geq \zeta^2+\frac{1}{\zeta^2}
\label{u-v}
\end{equation}
where $\hat{u}=|\zeta|\hat{x}_a+\frac{1}{\zeta}\hat{x}_b$ and
$\hat{v}=|\zeta|\hat{p}_a-\frac{1}{\zeta}\hat{p}_b$.
Our task is to find local unitary transformations which
remove all forms of nonclassicality other than entanglement. This
goal we approach by attempting to minimize the violation of
inequality (\ref{u-v}), in order to remove states which are
nonclassical, but separable, from the picture. We start by
rewriting the inequality (\ref{u-v}) as
\begin{eqnarray}
&&\zeta^2(\langle\hat{x}_a^2\rangle+\langle\hat{p}_a^2\rangle-1)+ \zeta^{-2}(\langle\hat{x}_b^2\rangle+\langle\hat{p}_b^2\rangle-1)\nonumber \\
&&+2\mbox{sign}(\zeta)(\langle\hat{x}_a\hat{x}_b\rangle-\langle\hat{p}_a\hat{p}_b\rangle)\geq
0. \label{u-v1}
\end{eqnarray}
Before proceeding further, let us apply preliminary local unitary
operations aimed at focusing all the information about the states
into moments featuring in Eq.~(\ref{u-v}), setting values of all
the remaining moments to zero, and transforming the subsystems
into classical \cite{simon}. Without loss of generality, we
can assume
$\langle\hat{x}_a\hat{x}_b\rangle\geq\langle\hat{p}_a\hat{p}_b\rangle$
and find that the left-hand side of Eq.~(\ref{u-v}) becomes
minimized when
\begin{equation}
\zeta^2=\frac{\sqrt{\langle\hat{x}_b^2\rangle+\langle\hat{p}_b^2\rangle-1}}{\sqrt{\langle\hat{x}_a^2\rangle+\langle\hat{p}_a^2\rangle-1}}~~;~~
\mbox{sign}(\zeta)=-1. \label{u-v2}
\end{equation}
Substituting these into Eq.~(\ref{u-v}), we find that the
inequality becomes
\begin{equation}
\prod_{k=a,b}(\langle\hat{x}_k^2\rangle+\langle\hat{p}_k^2\rangle-1)^{1/2}
\geq
(\langle\hat{x}_a\hat{x}_b\rangle-\langle\hat{p}_a\hat{p}_b\rangle)
\label{u-v3}
\end{equation}
Let us then apply local squeezing operations,
$\hat{x}_k\rightarrow \sqrt{s_k}\hat{x}_k\equiv\hat{x}_k^\prime$
and $\hat{p}_k\rightarrow
\hat{p}_k/\sqrt{s_k}\equiv\hat{p}_k^\prime$. Using the Lagrangian
multiplier to keep the left-hand side of the inequality invariant,
we find that squeezing operation which maximizes the right-hand
side of the inequality results in the following condition:
\begin{equation}
\left(\langle\hat{x}_a^{\prime^2}\rangle-\frac{1}{2}\right)
\left(\langle\hat{p}_b^{\prime^2}\rangle-\frac{1}{2}\right) =
\left(\langle\hat{x}_b^{\prime^2}\rangle-\frac{1}{2}\right)
\left(\langle\hat{p}_a^{\prime^2}\rangle-\frac{1}{2}\right)
\label{u-v4}
\end{equation}
under which the sufficient condition for entanglement (\ref{u-v})
then reads as
\begin{eqnarray}
&&\prod_{k = a,b}\left(\langle\hat{x}_k^{\prime^2}\rangle-\frac{1}{2}\right)^{1/2}
+ \prod_{k=1,2}\left(\langle\hat{p}_k^{\prime^2}\rangle-\frac{1}{2}\right)^{1/2}
\nonumber \\
&&~~~\geq |\langle\hat{x}_a^{\prime}\hat{x}_b^{\prime}\rangle| +
|\langle\hat{p}_a^{\prime}\hat{p}_b^{\prime}\rangle|. \label{u-v5}
\end{eqnarray}

On the other hand, it is straightforward to find that a positive
$P$ function is assigned to a Gaussian function when
\begin{eqnarray}
&&
\prod_{k=a,b}\left(\langle\hat{x}_k^{\prime^2}\rangle-\frac{1}{2}\right)^{1/2} \geq
|\langle\hat{x}_a^{\prime}\hat{x}_b^{\prime}\rangle| ~~~\mbox{and}
\nonumber \\
&&
\prod_{k=a,b}\left(\langle\hat{p}_k^{\prime^2}\rangle-\frac{1}{2}\right)^{1/2}
 \geq |\langle\hat{p}_a^{\prime}\hat{p}_b^{\prime}\rangle|. \label{u-v6}
\end{eqnarray}
Thus the two conditions (\ref{u-v5}) and (\ref{u-v6}) coincide if
\begin{eqnarray}
&&
\prod_{k = a,b}\left(\langle\hat{x}_i^{\prime^2}\rangle-\frac{1}{2}\right)^{1/2}
- \prod_{k=a,b}\left(\langle\hat{p}_i^{\prime^2}\rangle-\frac{1}{2}\right)^{1/2}
\nonumber \\
&&~~~= |\langle\hat{x}_a^{\prime}\hat{x}_b^{\prime}\rangle| -
|\langle\hat{p}_a^{\prime}\hat{p}_b^{\prime}\rangle|. \label{u-v7}
\end{eqnarray}
We conclude that when the two conditions (\ref{u-v4}) and
(\ref{u-v7}) are satisfied by local unitary transformations, the
existence of a positive $P$ function becomes the sufficient and
necessary condition for the separability of any Gaussian two-mode
field. For the special set of states which in the transformed form display $|\langle \hat{x}'_a\hat{x}'_b\rangle| = |\langle \hat{p}'_a\hat{p}'_b\rangle|$ these conditions coincide with those of refs. \cite{R1,R2,R4}. Interestingly enough, the complete condition is equivalent to the local unitary transformations found by Duan {\it et al.}~\cite{duan}.

\section{nonclassicality depth}
As was demonstrated in the previous section, a qualitative relation between nonclassicality and entanglement can be found between all Gaussian states. It might be pondered now, whether there is a quantitative relation as well. In \cite{R5,R3} it was shown that any measure of nonclassicality can be used as a measure of entanglement after appropriate \emph{nonlocal} operations have been performed. We now pose a different question. For a quantum state in the form found in Sec. III, is there a direct qualitative correspondence between nonclassicality and entanglement?

To shine some light on the issue, let us compare a measure of
entanglement, the logarithmic negativity~\cite{vidal}, to the depth of
nonclassicality \cite{leetx}, which was introduced as an amount
of thermal noise that is required to transform a quantum state into a
classical one. Formally, it is the smallest $T$ for which the
function
\begin{equation}
 P'(\alpha) = \int P(\beta) e^{-|\alpha-\beta|^2/T} d^2 \beta
\end{equation}
is regular and positive.

For simplicity, we introduce shorthand notations
for variances as follows:
\begin{eqnarray}
m_1=\langle\hat{x}_a^{\prime^2}\rangle~,~m_2=\langle\hat{p}_a^{\prime^2}\rangle~,~n_1=\langle\hat{x}_b^{\prime^2}\rangle \nonumber \\
n_2=\langle\hat{p}_b^{\prime^2}\rangle~,~c_1=\langle\hat{x}_a^{\prime}\hat{x}_b^\prime\rangle~,~c_2=\langle\hat{p}_a^{\prime}\hat{p}_b^\prime\rangle,
\label{new1}
\end{eqnarray}
while assuming all other moments featured in the covariance
matrix (See \cite{simon} for its definition) to be zero. The logarithmic negativity of such a state can
be expressed as $LN = -\log_2 (\sqrt{2 e})$ with
\begin{equation}
 e =  \Delta - \sqrt{\Delta^2 - 4 |\Sigma|}, \quad \Delta = m_1 m_2 + n_1 n_2 - 2 c_1 c_2,
\end{equation}
where $|\Sigma|$ denotes the determinant of the covariance
matrix, which, under our assumed conditions can be expressed as
$|\Sigma|= (m_1n_1-c_1 c_2)(m_2n_2-c_1 c_2)$ \cite{adesso}. The other quantity,
the depth of nonclassicality is
\begin{equation}
T = \min_{k = 1,2}~\left[\frac{1}{2}(1-m_k - n_k + \sqrt{4 c_k^2 +(m_k -n_k)^2})\right] .
\end{equation}
Let us start the comparison by considering a specific scenario --
an entangled state for which $c_1 = -c_2 = c$. The conditions
(\ref{u-v4}) and (\ref{u-v7}) in turn require that $m_1 = m_2 = m$
and $n_1 = n_2 = n$, thus allowing the nonclassicality
depth to be defined without the need for minimization and
reducing the expression relevant for logarithmic negativity
to
\begin{equation}
e = m^2 + n^2 - 2c^2 - (m+n)\sqrt{4 c^2 +(m-n)^2}.
\end{equation}
If we assume that there is no strict relation between
nonclassicality and entanglement, there should exist at least two
different states with the same depth of nonclassicality but
different values of logarithmic negativity.  In other words, for
the two sets of parameters $m$, $n$ and $c$ and $m' = m+\delta m$,
$n' = n+\delta n$ and $c' = c+\delta c$, the depths of
nonclassicality should be equal, $T' = T$, while the entanglement
should differ by
\begin{eqnarray}\label{ediff}
& &\Delta e = e' - e
= 2 (\delta c^2 + 2 c\delta c -
\nonumber \\
&&~~~~~~~~~ \delta m \delta n - m \delta n - n \delta m) + \chi (\delta m + \delta n),
\end{eqnarray}
where $\chi = m+n -\sqrt{4 c^2 - (m-n)^2}$. From the equality of the nonclassicality depth,
\begin{eqnarray}
\delta m + \delta n &=&  \sqrt{4(c+\delta c)^2 + (m-n + \delta m - \delta n)^2} \nonumber \\
& & - \sqrt{4c^2 + (m-n)^2}.
\end{eqnarray}
This in conjunction with
\begin{eqnarray}
& &\frac{4 \delta c^2+ 8 c \delta c + (\delta m-\delta n)^2 + 2(m-n)(\delta m - \delta n)}{\delta m + \delta n} \nonumber \\
&=& \sqrt{4(c+\delta c)^2 + (m-n + \delta m - \delta n)^2} \nonumber \\
& &+ \sqrt{4c^2 + (m-n)^2}
\end{eqnarray}
leads to
\begin{eqnarray}
\frac{(\delta m + \delta n)\chi}{2}
=  \delta m \delta n + m \delta n +  n \delta m - 2 c \delta c -  \delta c^2.
\end{eqnarray}
Substituting this expression into Eq. (\ref{ediff}) we arrive at
$\Delta e = 0$. So even though there may be two different states
with the same depth of nonclassicality, their entanglement,
measured by logarithmic negativity, is also the same. In other
words, surprisingly there is a one-to-one correspondence between
logarithmic negativity and entanglement depth. This suggests that in the particular form of the quantum state, all of the nonclassicality is in the form of entanglement.

\begin{figure}
\centerline{\psfig{figure=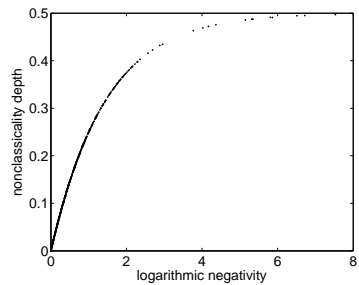,width=0.6\linewidth}}
\caption{Relation between the depth of nonclassicality and the
logarithmic negativity \petr{for states satisfying conditions (\ref{u-v4}) and (\ref{u-v7})} - Monte Carlo simulation.} \label{LNdepth}
\end{figure}

It is still uncertain whether this conjecture holds for all
Gaussian states which satisfy conditions (\ref{u-v4}) and
(\ref{u-v7}) but a Monte Carlo simulation , as shown in
Fig.~\ref{LNdepth}, suggests it may be so.

\section{Remarks}
Entanglement is a manifestation of nonclassicality of multipartite states. In this regard, entangled states are a subset of nonclassical states. This can be easily understood by recognizing that since nonclassicality was originally defined for single mode states, local unitary operation, which does not change entanglement, can alter nonclassicality of the state significantly. However, an equivalency between entanglement and nonclassicality might be found if local unitary transformation, aimed at reduction of the single-mode nonclassicality, were considered in conjunction with the respective criteria. Intuitively, one could reason that after transformation, which leaves local subsystems of the state classical, any sign of nonclassicality is a mark of entanglement. This condition is, however, too weak, as was shown in Sec. II.

For Gaussian states, one can find explicit local unitary operations, which minimize the nonclassicality of the state. After this transformation, criteria of nonclassicality and entanglement are equivalent, and this equivalence holds for all Gaussian states. In contrast, the conditions of equivalence presented in \cite{R1,R2,R4} require a special kind of symmetric state.

We have also shown the equivalency between the depth of nonclassicality and the logarithmic negativity for a special case of states with isotropic local quadrature variances. However, our numerical analysis suggests that this relation may hold for all Gaussian states after necessary local unitary transformations have been performed. Proving this conjecture remains an open problem.

\acknowledgements We thank Dr. M. Paternostro for stimulating
discussions and acknowledge financial support from the UK EPSRC,
the QIP IRC and the European Social Fund.
P.M acknowledges financial support from the Ministry of
Education of the Czech Republic (Grants No. LC06007
and No. MSM6198959213), and of the Future and Emerging Technologies (FET)
programme within the Seventh Framework Programme
for Research of the European Commission, under the
FET-Open grant agreement COMPAS, number 212008.

\end{document}